\documentclass[12pt]{article}
\newcommand{\interlinia}{}
\title{Flexoelectric polarization
in the biaxial nematic phase}
\author{A. Kapanowski\\
{\em Marian Smoluchowski Institute of Physics,}\\
{\em Jagellonian University, ulica Reymonta 4,}\\
{\em 30-059 Krak\'{o}w, Poland}  }
\begin{document}
\maketitle

\begin{abstract}
\interlinia
The dipole flexoelectric (FE) polarization
in liquid crystals is derived in the thermodynamic limit
at small distortions and small density.
General microscopic expressions for the FE
coefficients are obtained in the case of the uniaxial
and biaxial nematic phases composed of $C_{2v}$ molecules.
The expressions involve the one-particle distribution function
and the potential energy of two-body short-range interactions.
In the case of the biaxial nematic phase, six basic deformations
produce FE polarization but there are only five independent
FE coefficients.
\newline\newline
PACS number(s): 61.30.Cz, 77.84.Nh
\end{abstract}

\interlinia

Biaxial nematic liquid crystals are characterized by anisotropic
positional short-range order and orientational long-range order
\cite{[1993_de_Gennes]}.
The anisotropic molecules tend to be parallel to selected axes,
labeled by the unit orthogonal vectors $\vec{L}$, $\vec{M}$, 
and $\vec{N}$.
In the uniaxial nematic phase
only the $\vec{N}$ axis is defined.
Stable biaxial phases were observed in 1980 in lyotropic systems
by Yu and Saupe 
\cite{[1980_Yu_Saupe]}
but their existence in thermotropic systems was not certain
for many years
\cite{[2000_Berardi_Zannoni]}.
Recently Madsen {\em et al.} 
\cite{[2004_Madsen]}
synthesized banana-shaped mesogens
and evidence for biaxiality was achieved using NMP spectroscopy.
Acharya {\em et al.} 
\cite{[2004_Acharya]}
revealed biaxiality of bent-core mesogens by means of
low-angle x-ray diffraction.
Merkel {\em et al.}
\cite{[2004_Merkel]}
carried out infrared absorbance measurements on two liquid crystalline
organo-siloxane tetrapodes and showed the existence of a biaxial nematic
phase.

Theoretical predictions of biaxial nematic phases started 
in 1970 with the paper by Freiser
\cite{[1970_Freiser]}.
Then they were studied using a number of theoretical methods, e.g. 
mean-field theory 
\cite{[1974_Straley]},
\cite{[1982_Mulder_Ruijgrok]},
counting methods 
\cite{[1972_Shih_Alben]},
\cite{[1994_Li_Freed]},
Landau-de Gennes theory 
\cite{[1973_Alben]},
\cite{[1986_Gramsbergen_Longa_Jeu]},
bifurcation analysis 
\cite{[1989_Mulder]},
and density-functional theory 
\cite{[1990_Holyst_Poniewierski]}.
All the theories mentioned above predict that the system will
exhibit four phases as the molecular biaxiality varies:
a positive and a negative uniaxial phases, respectively, formed by prolate 
and oblate molecules, a biaxial, and an isotropic phase.
The nwmatic-isotropic phase transition is expected to be first order
and to weaken as the biaxiality increases until it becomes continuous at 
the point (Landau bicritical point) of maximum molecular biaxiality.
At this point the system should go directly from a biaxial to an isotropic 
phase. The uniaxial-biaxial transition is expected to be second order.

A microscopic mean field theory
\cite{[2003_Sonnet_Virga_Durand]},
\cite{[2006_Bisi]}
predicts the possibility of lines of second-order and first-order
uniaxial-biaxial transitions joined at a tricritical point.
The experimental results by Merkel {\em et al.}
\cite{[2004_Merkel]}
were successfully interpreted in terms of this theory.
A weakly first-order uniaxial-biaxial transition was also revealed
by dynamic light scattering from orientational order fluctuations 
in a liquid crystalline tetrapode
\cite{[2006_Neupane]}.
Recently, two-particle cluster theory was applied to study the biaxial
molecules in the Sonnet model 
\cite{[2003_Sonnet_Virga_Durand]}
and qualitative agreement with the mean-field theory was obtained
\cite{2006_Zhang}.
A Monte Carlo study of biaxial nematic phases composed of V-shaped
molecules was done by Bates and Luckhurst
\cite{[2005_Bates_Luckhurst]}.

In most practical circumstances the liquid crystal phase alignment
is deformed.
The deformations usually are described by a continuum theory
where the free energy is expanded to the second order around
the free energy of the undeformed state in ascending powers
of a parameter that measures the deformation.
The free energy due to the distortion of the axes is expressed
in terms of the vector spatial derivatives and the elastic constants.

In a deformed uniaxial nematic liquid crystal, there should appear
in many cases a spontaneous dielectric polarization
described by Meyer \cite{[1969_Meyer]}:
\begin{equation}
\label{Meyer-P}
\vec{P} = e_1 \vec{N} (\vec{\nabla} \cdot \vec{N})
+ e_3 (\vec{N} \cdot \vec{\nabla}) \vec{N}
= \mbox{} e_1 \vec{N} (\vec{\nabla} \cdot \vec{N})
- e_3 \vec{N} \times (\vec{\nabla} \times \vec{N} ),
\end{equation}
where $e_1$ and $e_3$ are the splay and the bend flexoelectric (FE)
coefficients, respectively.
The appearance of spontaneous polarization in liquid crystals
as a result of orientational deformations is referred to as
the flexoelectric effect.

A microscopic mechanism of the FE effect was proposed by Meyer
\cite{[1969_Meyer]}, 
who pointed out that under
the condition of orientation deformation of a liquid crystal,
banana-shaped or conical molecules should be so oriented
that their constant dipoles are ordered and macroscopic
polarization sets in.
On the other hand, Prost and Marcerou
\cite{[1977_Prost_Marcerou]}
showed that polarization in a deformed liquid crystal 
is also produced as a result
of a gradient in the average density of the molecule quadrupole moments. 
Such a polarization does not need asymmetry
in molecular shape of the banana or cone kind.
Later, the FE coefficients for the uniaxial nematic phase
were calculated by means of a statistical-mechanical theory
\cite{[1976_Straley]},
mean-field theories
\cite{[1971_Helfrich]},
\cite{[1971_Derzhanski_Petrov]},
the density-functional formalism
\cite{[1989_Singh_Singh]},
\cite{[1991_Somoza_Tarazona]},
an integral equation approach
\cite{[2001_Zakharov_Dong]},
and computer simulations
\cite{[1999_Stelzer_Berardini_Zannoni]},
\cite{[2000_Billeter_Pelcovits]}.
The expressions connecting the molecular asymmetry, the elasticity
of the liquid crystal, and the FE coefficients were obtained
by Derzhanski and co-workers
\cite{[1981_Derzhanski]}.
It was also shown
\cite{[2001_Ferrarini]}
that the FE coefficients can have nontrivial dependence
on the details of the molecule's chemical structure
(an isomerization, a charge distribution).

The flexoelectric polarization can influence electrooptical properties,
defect formation, and structural instability.
Thus, different techniques have been suggested to observe
possible mechanisms producing the FE effect
\cite{[2001_Petrov]}.
The FE coefficients can be obtained experimentally from
measuring the polarizations or the surface charges
induced by an imposed distortion or using the inverse effect,
because when an electric field is applied on a nematic, the alignment
may become distorted and this will imply a polarization
\cite{[1972_Schmidt_Schadt_Helfrich]}.
The flexoelectric polarization of 5CB
was measured by means of a pyroelectric-effect-based technique
\cite{[2000_Blinov_et_al]},
\cite{[2001_Blinov_et_al]}
and recently a technique inspired by the flexoelectric-optic effect
was demonstrated
\cite{[2006_Ewings_et_al]}.

Below, we would like to investigate the FE effect 
in biaxial nematic phases.
In the case of the uniaxial nematic phase there are three
independent phase deformations: splay, twist, and bend.
The symmetry considerations of Rudquist and Lagerwall
\cite{[1997_Rudquist_Lagerwall]}
prove that in uniaxial nematic liquid crystals the FE effect
can be induced by splay or bend.
The polarization connected with bend has to be perpendicular 
to the director $\vec{N}$, whereas in the splay deformation,
a polarization along the director is admitted.
The twist is not connected with a local polarization of the medium
because there is always a two-fold symmetry axis perpendicular
to the helix axis.

It was shown \cite{[1997_Kapanowski]}
that in the case of the biaxial nematic phase
there are 12 independent phase deformations and
six of them are connected with splays and bends
of the vectors $\vec{L}$, $\vec{M}$, and $\vec{N}$.
Thus, we can expect that there are six FE coefficients
for the biaxial nematic phase.

Let us assume that the FE polarization, which is a vector quantity,
depends on the spatial derivatives of the vectors
$\vec{L}$, $\vec{M}$, and $\vec{N}$
\begin{equation}
\label{Pmacro}
P_{\alpha} = s_{ij} R_{i\alpha} \partial_{\beta} R_{j\beta}
+ b_{ij} R_{i\beta} \partial_{\beta} R_{j\alpha},
\end{equation}
where we denoted
\begin{equation}
\label{LMN=R}
L_{\alpha} = R_{1\alpha},
\ M_{\alpha} = R_{2\alpha},
\ N_{\alpha} = R_{3\alpha}.
\end{equation}
The matrix elements $R_{i\alpha}$  
($i=1, 2, 3$ and $\alpha=x, y, z$) satisfy the conditions
\begin{equation}
\sum_{\alpha} R_{i\alpha} R_{j\alpha} = \delta_{ij},
\ \sum_{i} R_{i\alpha} R_{i\beta} = \delta_{\alpha\beta}.
\end{equation}
The above relations express the orthogonality and the completeness 
of the local frame. It is also possible to derive the identity
(to be used later)
\begin{equation}
\label{tozsamosc}
\sum_{i} ( R_{i\alpha} \partial_{\beta} R_{i\beta} 
+ R_{i\beta} \partial_{\beta} R_{i\alpha})= 0.
\end{equation}

The number of independent FE coefficients will be determined
with the help of symmetry requirements.
The local frame can be transformed into the new one
\begin{equation}
R_{i\alpha}' = T_{ij} R_{j\alpha},
\end{equation}
where $T_{ij}$ ($i,j=1,2,3$) are the elements 
of the orthogonal transformation.
The polarization may be expressed in terms of new variables
with new (primed) FE coefficients.
As the FE coefficients do not change under symmetry operations,
we can identify the independent FE coefficients.
As a result we get six FE coefficients
$s_{ii}$ and $b_{ii}$.
But from the identity (\ref{tozsamosc}) we conclude that
the coefficients $s_{ii}$ and $b_{ii}$ are not unique.
The same polarization $P$ can be obtained by means of the transformed
coefficients
\begin{equation}
s'_{ii} = s_{ii}+c,\ b'_{ii} = b_{ii}+c,
\end{equation}
where $c$ is any constant. Thus, only five coefficients are independent.
Note that the differences $s_{ii}-s_{jj}$, $b_{ii}-b_{jj}$, 
or $s_{ii}-b_{jj}$ do not depend on the constant $c$.

Now we consider six small deformations of the directors 
($\vec{L},\vec{M},\vec{N}$) of the form
\cite{[1997_Kapanowski]}
\begin{eqnarray}
\vec{L}(\vec{r}) & = & [1,q_7 y-q_5 x, q_9 z-q_4 x], 
\nonumber \\
\vec{M}(\vec{r}) & = & [q_5 x-q_7 y, 1, q_8 z-q_6 y], 
\nonumber \\
\label{LMN(qi)}
\vec{N}(\vec{r}) & = & [q_4 x-q_9 z, q_6 y-q_8 z, 1],
\end{eqnarray}
where the parameters $q_i$ ($i=4,\ldots,9$) 
describe the deformations
($1/q_i$ is a certain length much greater than the size of the sample).
The corresponding FE polarization has the form
\begin{eqnarray}
& P_x = q_7 (s_{11}-b_{22}) + q_9 (s_{11}-b_{33}) = q_7 a_7 + q_9 a_9, &
\nonumber \\
& P_y = q_5 (s_{22}-b_{11}) + q_8 (s_{22}-b_{33}) = q_5 a_5 + q_8 a_8, &
\nonumber \\
& P_z = q_4 (s_{33}-b_{11}) + q_6 (s_{33}-b_{22}) = q_4 a_4 + q_6 a_6, &
\label{P(qi)}
\end{eqnarray}
where we introduced six physical FE coefficients $a_i$ which
satisfy the identity
\begin{equation}
a_4 - a_5 - a_6 + a_7 + a_8 - a_9 = 0.
\end{equation}
Eqs. (\ref{P(qi)}) are phenomenological expressions
and we should provide alternative microscopic expressions
in order to obtain microscopic expressions for the FE
coefficients.

Let us consider a system of $N$ rigid molecules with the 
$C_{2v}$ symmetry. Such molecules can form uniaxial 
($D_{\infty h}$) and biaxial ($D_{2h}$) nematic phases.
The free energy for the system can be derived 
in the thermodynamic limit 
($N\rightarrow\infty$, $V\rightarrow\infty$, $N/V=\mbox{const}$) 
from the Born-Bogoliubov-Green-Kirkwood-Yvon hierarchy 
\cite{[1980_Reichl]}
or as a cluster expansion 
\cite{[1979_Stecki_Kloczkowski]}.
The total free energy $F$ consists of the entropy term
and the interaction term
\begin{equation}
\label{Ftotal}
F = F_{ent} + F_{int},
\end{equation}
where
\begin{eqnarray}
\beta F_{ent} &=& 
\int {d\vec{r}}{dR} G(\vec{r},R)
 \{ \ln [ G(\vec{r},R) \Lambda ]-1 \},
\\
\beta F_{int} &=& 
- {\frac {1}{2}} \int {d\vec{r}_1}{dR_1}{d\vec{r}_2}{dR_2} 
G(\vec{r}_1,R_1) G(\vec{r}_2,R_2) f_{12}.
\end{eqnarray}
We define
$dR = d\phi d\theta \sin\theta d\psi$, 
$f_{12}=\exp (-\beta \Phi_{12})-1 $ (the Mayer function), 
$\beta = 1/(k_{B} T)$, 
$\vec{u}=\vec{r}_2-\vec{r}_1=u\vec{\Delta}$;
and $\Lambda$ is related to the ideal gas properties.
The normalization of the one-particle distribution function $G$ is
\begin{equation}
\label{normaG}
\int {d\vec{r}}{dR} G(\vec{r},R)=N. 
\end{equation}
The equilibrium distribution $G$ minimizing the free energy 
(\ref{Ftotal}) satisfies the equation
\begin{equation}
\label{lnGgeneral}
\ln [G(\vec{r}_1,R_1) \Lambda]
- \int {d\vec{r}_2}{dR_2} G(\vec{r}_2,R_2) f_{12} = \mbox{const}.
\end{equation}
In the homogeneous biaxial nematic phase
composed of $C_{2v}$ or $D_{2h}$ molecules
the distribution function has the form
\cite{[1997_Kapanowski]}
\begin{equation}
G_0(R) = 
G_0(\vec{l} \cdot \vec{L}, \vec{l} \cdot \vec{N},
\vec{n} \cdot \vec{L}, \vec{n} \cdot \vec{N}),
\end{equation}
where the unit orthogonal vectors $(\vec{l}, \vec{m}, \vec{n})$ 
describe the molecule's orientation.
For the $C_{2v}$ molecules, the molecule symmetry axis
is determined by the vector $\vec{n}$. 
Note that this is the {\em long} axis of wedge-shaped molecules
and the {\em short} axis of banana-shaped molecules.
In order to derive expressions for the elastic constants it is
enough to assume that, in the deformed phase,
the phase orientation depends on the position
but the magnitude of the alignment is constant,
\begin{equation}
G_0(\vec{r},R) = 
G_0[\vec{l} \cdot \vec{L}(\vec{r}), 
\vec{l} \cdot \vec{N}(\vec{r}),
\vec{n} \cdot \vec{L}(\vec{r}), 
\vec{n} \cdot \vec{N}(\vec{r})].
\end{equation}
In order to derive expressions for the FE coefficients
we have to take into account a small change
of the alignment,
\begin{equation}
\label{G=G0(1+g)}
G(\vec{r},R) = G_0(\vec{r},R)[1+g(\vec{r},R)],
\end{equation}
where $g$ is expected to be small.
The microscopic polarization depends on the position
inside the phase via the distribution function
\begin{equation}
\label{Pmicro}
\vec{P}(\vec{r}) = \int {dR} G(\vec{r},R) \vec{\mu}(R). 
\end{equation}
The molecule electric dipole moment can be defined
in the molecular frame as 
\begin{equation}
\mu_{\alpha}(R) = 
\mu_1 l_{\alpha} + \mu_2 m_{\alpha} + \mu_3 n_{\alpha},
\ \alpha = x,y,z. 
\end{equation}
According to Straley \cite{[1976_Straley]}, 
the function $g$ can be obtained from the expression
\begin{equation}
\label{g-general}
g(\vec{r},R_1) = \int {d\vec{u}} {dR_2} f_{12} 
(\vec{u} \cdot \vec{\nabla}) G_0(\vec{r},R_2).
\end{equation}
Finally, from Eqs. (\ref{Pmicro}), (\ref{g-general}), and
(\ref{LMN(qi)}), we get the components of the 
microscopic FE polarization,
\begin{eqnarray}
P_x & = & \int {d\vec{u}}{dR_1}{dR_2} f_{12} G_0(R_1) \mu_{x}(R_1)
[(U_{2z}-W_{2x})q_9 u_z + U_{2y} q_7 u_y],
\nonumber \\
P_y & = & \int {d\vec{u}}{dR_1}{dR_2} f_{12} G_0(R_1) \mu_{y}(R_1)
[W_{2y}(-q_8 u_z) + U_{2y}(-q_5 u_x)],
\nonumber \\
P_z & = & \int {d\vec{u}}{dR_1}{dR_2} f_{12} G_0(R_1) \mu_{z}(R_1)
[(U_{2z}-W_{2x})(-q_4 u_x) + W_{2y} q_6 u_y],
\label{P(qi)micro}
\end{eqnarray}
where we denoted
\begin{equation}
U_{\alpha} =\partial_{1} G_{0} l_{\alpha}+\partial_{3} G_{0} n_{\alpha},
\ W_{\alpha} = \partial_{2} G_{0} l_{\alpha}+\partial_{4} G_{0} n_{\alpha}.
\end{equation}
When we compare Eqs. (\ref{P(qi)}) and (\ref{P(qi)micro}),
we obtain the equations for the FE coefficients:
\begin{eqnarray}
a_4 & = & \int {d\vec{u}}{dR_1}{dR_2} f_{12} G_0(R_1)
\mu_{z}(R_1) (-u_x) (U_{2z}-W_{2x}),
\nonumber \\
a_5 & = & \int {d\vec{u}}{dR_1}{dR_2} f_{12} G_0(R_1)
\mu_{y}(R_1) (-u_x)U_{2y},
\nonumber \\
a_6 & = & \int {d\vec{u}}{dR_1}{dR_2} f_{12} G_0(R_1)
\mu_{z}(R_1) u_y W_{2y},
\nonumber \\
a_7 & = & \int {d\vec{u}}{dR_1}{dR_2} f_{12} G_0(R_1)
\mu_{x}(R_1) u_y U_{2y},
\nonumber \\
a_8 & = & \int {d\vec{u}}{dR_1}{dR_2} f_{12} G_0(R_1)
\mu_{y}(R_1) (-u_z) W_{2y},
\nonumber \\
a_9 & = & \int {d\vec{u}}{dR_1}{dR_2} f_{12} G_0(R_1)
\mu_{x}(R_1) u_z (U_{2z}-W_{2x}).
\end{eqnarray}
Let us consider the uniaxial nematic phase composed of
wedge-shaped molecules. The long molecule axes $\vec{n}$ 
are almost parallel to the $\vec{N}$ axis, and we get
\begin{equation}
U_{\alpha}=0,\ a_5 = a_7 = 0.
\end{equation}
We recover the Meyer expression (\ref{Meyer-P})
with two FE coefficients,
\begin{equation}
\label{stozek}
e_1 = a_4 = a_6,\ e_3 = -a_8 = -a_9.
\end{equation}
Finally we consider the uniaxial nematic phase composed of
banana-shaped molecules. The long molecule axes $\vec{l}$
are almost parallel to the $\vec{L}$ axis, and we get
\begin{equation}
W_{\alpha}=0,\ a_6 = a_8 = 0.
\end{equation}
We obtain the Meyer expression (\ref{Meyer-P}),
where $\vec{N}$ is replaced with $\vec{L}$.
The FE coefficients are
\begin{equation}
\label{banan}
e_1 = a_7 = a_9,\ e_3 = -a_4 = -a_5.
\end{equation}
Note that Eqs. (\ref{stozek}) and (\ref{banan})
describe $C_{2v}$ molecules, whereas very often
simpler expressions for $C_{\infty v}$ molecules are present
in the literature.

In conclusion, we derived the microscopic formulas for the six FE
coefficients in the case of the biaxial nematic phase
with the $C_{2 v}$ molecules. It appeares that only five 
FE coefficients are independent.
In order to calculate the values of the FE coefficients one needs
the one-particle distribution function and the potential energy
of molecular interactions.
The Meyer expressions \cite{[1969_Meyer]}
are recovered in the case of the wedge-shaped
and banana-shaped molecules in the uniaxial nematic phase.
Generally, the splitting of the two Meyer FE coefficients 
and the appearance of new small FE coefficients
are expected at the uniaxial-biaxial nematic transition.
In order to describe real substances,
the presented results should be generalized 
beyond the low-density limit, where the Mayer function $f_{12}$
is replaced with a better approximation of the direct
correlation function $c_2$. 
On the other hand, other sources of dielectric polarization
should be taken into account: the quadrupole contribution
or the ordoelectric polarization.
But even then, a qualitative comparison
between theory and experiment may be difficult, because
the experimental data on FE coefficients are still scarce and
sometimes contradictory \cite{[2001_Petrov]}.

\section*{ACKNOWLEDGMENTS}

This work was supported by the Faculty of Physics, Astronomy and
Applied Computer Science, Jagellonian University (WRBW Grant No. 45/06).
The author is also grateful to J. Spa{\l}ek for his support.


\newpage

\begin{thebibliography}{99}

\bibitem{[1993_de_Gennes]}
P. G. de Gennes and J. Prost, {\em The Physics of Liquid Crystals}
(Clarendon Press, Oxford, 1993).

\bibitem{[1980_Yu_Saupe]}
L. J. Yu, A. Saupe, Phys. Rev. Lett. {\bf 45}, 1000 (1980).

\bibitem{[2000_Berardi_Zannoni]}
R. Berardi and C. Zannoni, J. Chem. Phys. {\bf 113}, 5971 (2000).

\bibitem{[2004_Madsen]}
L. A. Madsen, T. J. Dingemans, M. Nakata, and E. T. Samulski,
Phys. Rev. Lett. {\bf 92}, 145505 (2004).

\bibitem{[2004_Acharya]}
B. R. Acharya, A. Primak, and S. Kumar, 
Phys. Rev. Lett. {\bf 92}, 145506 (2004).

\bibitem{[2004_Merkel]}
K. Merkel, A. Kocot, J. K. Vij, R. Korlacki, G. H. Mehl,
and T. Meyer, Phys. Rev. Lett. {\bf 93}, 237801 (2004).


\bibitem{[1970_Freiser]}
M. J. Freiser, Phys. Rev. Lett. {\bf 24}, 1041 (1970).

\bibitem{[1974_Straley]}
J. P. Straley, Phys. Rev. A {\bf 10}, 1881 (1974).

\bibitem{[1982_Mulder_Ruijgrok]}
B. M. Mulder, Th. W. Ruijgrok, Physica A {\bf 113}, 145 (1982).

\bibitem{[1972_Shih_Alben]}
C.-S. Shih and R. Alben, J. Chem. Phys. {\bf 57}, 3055 (1972).

\bibitem{[1994_Li_Freed]}
W. Li and K. F. Freed, J. Chem. Phys. {\bf 101}, 519 (1994).

\bibitem{[1973_Alben]}
R. Alben, Phys. Rev. Lett. {\bf 30}, 778 (1973).

\bibitem{[1986_Gramsbergen_Longa_Jeu]}
E. F. Gramsbergen, L. Longa, W. H. de Jeu, 
Phys. Rep. {\bf 135}, 195 (1986).

\bibitem{[1989_Mulder]}
B. Mulder, Phys. Rev A {\bf 39}, 360 (1989).

\bibitem{[1990_Holyst_Poniewierski]}
R. Holyst and A. Poniewierski, Mol. Phys. {\bf 69}, 193 (1990).


\bibitem{[2003_Sonnet_Virga_Durand]}
A. M. Sonnet, E. G. Virga, and G. E. Durand,
Phys. Rev. E {\bf 67}, 061701 (2003).

\bibitem{[2006_Bisi]}
F. Bisi, E. G. Virga, E. C. Gartland Jr., G. De Matteis, A. M. Sonnet,
and G. E. Durand, Phys. Rev. E {\bf 73}, 051709 (2006).

\bibitem{[2006_Neupane]}
K. Neupane, S. W. Kang, S. Sharma, D. Carney, T. Meyer, G. H. Mehl,
D. W. Allender, S. Kumar, and S. Sprunt, 
Phys. Rev. Lett. {\bf 97}, 207802 (2006).

\bibitem{2006_Zhang}
Zhi-Dong Zhang, Yan-Jun Zhang, and Zong-Li Sun,
Chin. Phys. Lett. {\bf 23}, 3025 (2006).

\bibitem{[2005_Bates_Luckhurst]}
M. A. Bates and G. R. Luckhurst, Phys. Rev. E {\bf 72}, 051702 (2005).


\bibitem{[1969_Meyer]}
R. B. Meyer, Phys. Rev. Lett. {\bf 22}, 918 (1969).

\bibitem{[1977_Prost_Marcerou]}
J. Prost and J. P. Marcerou, J. Phys. (Paris) {\bf 38}, 315 (1977).

\bibitem{[1976_Straley]}
J. P. Straley, Phys. Rev. A {\bf 14}, 1835 (1976).

\bibitem{[1971_Helfrich]}
W. Helfrich, Z. Naturforsch. A {\bf 26}, 833 (1971). 

\bibitem{[1971_Derzhanski_Petrov]}
A. Derzhanski and A. G. Petrov, Phys. Lett. A {\bf 36}, 427 (1971).

\bibitem{[1989_Singh_Singh]}
Y. Singh and U. P. Singh, Phys. Rev. A {\bf 39}, 4254 (1989).

\bibitem{[1991_Somoza_Tarazona]}
A. M. Somoza and P. Tarazona, Mol. Phys. {\bf 72}, 911 (1991).

\bibitem{[2001_Zakharov_Dong]}
A. V. Zakharov and R. Y. Dong, Eur. Phys. J. E {\bf 6}, 3 (2001).

\bibitem{[1999_Stelzer_Berardini_Zannoni]}
J. Stelzer, R. Berardini and C. Zannoni,
Chem. Phys. Lett. {\bf 299}, 9 (1999).

\bibitem{[2000_Billeter_Pelcovits]}
J. L. Billeter and R. A. Pelcovits, Liq. Cryst. {\bf 27}, 1151 (2000).

\bibitem{[1981_Derzhanski]}
A. Derzhanski, A. G. Petrov, and I. Bivas,
in {\em Advances in Liquid Crystals Research and Applications},
edited by L. Bata, (Pergamon Press, Akademiai Kiado,
Budapest, 1981), Vol. 1.

\bibitem{[2001_Ferrarini]}
A. Ferrarini, Phys. Rev. E {\bf 64}, 021710 (2001).

\bibitem{[2001_Petrov]}
A. G. Petrov, in {\em Physical Properties of Liquid Crystals, EMIS
Datareview Series 25}, edited by D. A. Dunmur, A. Fukuda, and 
G. L. Luckhurst (IEE, London, 2001).

\bibitem{[1972_Schmidt_Schadt_Helfrich]}
D. Schmidt, M. Schadt, and W. Helfrich, 
Z. Naturforsch. A {\bf 27}, 277 (1972).

\bibitem{[2000_Blinov_et_al]}
L. M. Blinov, M. I. Barnik, M. Ozaki, N. M. Shtykov, and
K. Yoshino, Phys. Rev. E {\bf 62}, 8091 (2000).

\bibitem{[2001_Blinov_et_al]}
L. M. Blinov, M. I. Barnik, H. Ohoka,  M. Ozaki, N. M. Shtykov, and
K. Yoshino, Eur. Phys. J. E {\bf 4}, 183 (2001).

\bibitem{[2006_Ewings_et_al]}
R. A. Ewings, C. Kischka, L. A. Parry-Jones, and S. J. Elston,
Phys. Rev. E {\bf 73}, 011713 (2006).


\bibitem{[1997_Rudquist_Lagerwall]}
P. Rudquist and S. T. Lagerwall, Liq. Cryst. {\bf 23}, 503 (1997).

\bibitem{[1997_Kapanowski]} 
A. Kapanowski, Phys. Rev. E {\bf 55}, 7090 (1997).

\bibitem{[1980_Reichl]}
L. E. Reichl, {\em A Modern Course in Statistical Physics}
(Edward Arnold Publishers, London, 1980).

\bibitem{[1979_Stecki_Kloczkowski]}
J. Stecki, A. Kloczkowski, J. Phys. Paris {\bf 40}, C3-360 (1979).

\end{thebibliography}
\end{document}